\documentclass[aps,prd,floats,nofootinbib]{revtex4}
\usepackage{epsfig}

\begin{document}

\title{Predictions for high energy neutrino cross-sections from the ZEUS
global PDF fits\\}
\author{Amanda Cooper-Sarkar}
\affiliation{Particle Physics, University of Oxford, Keble Road,
             Oxford OX1 3RQ, UK}
\author{Subir Sarkar}
\affiliation{Rudolf Peierls Centre for Theoretical Physics, 
             University of Oxford, 1 Keble Road, Oxford OX1 3NP, UK\\}
\date{\today}
\bigskip
\begin{abstract}
  We have updated predictions for high energy neutrino and
  antineutrino charged current cross-sections within the conventional
  DGLAP formalism of NLO QCD using a modern PDF fit to HERA data,
  which also accounts in a systematic way for PDF uncertainties
  deriving from both model uncertainties and from the experimental
  uncertainties of the input data sets. Furthermore the PDFs are
  determined using an improved treatment of heavy quark thresholds. A
  measurement of the neutrino cross-section much below these
  predictions would signal the need for extension of the conventional
  formalism as in BFKL resummation, or even gluon recombination
  effects as in the colour glass condensate model.
\end{abstract}

\pacs{xxx} 
\maketitle

\section{Introduction}
\label{sec:intro}

Predictions of neutrino cross-sections at high energies have sizeable
uncertainties which derive largely from the measurement uncertainties
on the parton distribution functions (PDFs) of the nucleon. In the
framework of the quark-parton model, high energy scattering accesses
very large values of $Q^2$, the invariant mass of the exchanged vector
boson, and very small values of Bjorken $x$, the fraction of the
momentum of the incoming nucleon taken by the struck quark. Thus when
evaluating uncertainties on high energy neutrino cross-sections it is
important to use the most up to date information from the experiments
at HERA, which have accessed the lowest-$x$ and highest $Q^2$ scales
to date. The present paper uses the formalism of the ZEUS-S global PDF
fits~\cite{Chekanov:2002pv}, updated to include {\em all} the HERA-I
data.

Conventional PDF fits use the Next-to-leading-order (NLO)
Dokshitzer-Gribov-Lipatov-Altarelli-Parisi (DGLAP)
formalism~\cite{Altarelli:1977zs,Gribov:1972ri,Lipatov:1974qm,Dokshitzer:1977sg}
of QCD to make predictions for deep inelastic scattering (DIS)
cross-sections of leptons on hadrons. At low-$x$ where the gluon
density is rising rapidly it is probably necessary to go beyond the
DGLAP formalism in order to sum $\ln(1/x)$ diagrams, as in the
Balitsky-Fadin-Kuraev-Lipatov (BFKL)
formalism~\cite{Kuraev:1977fs,Balitsky:1978ic,Lipatov:1985uk} (for
recent work
see~\cite{White:2006xv,Altarelli:2005ni,Ciafaloni:2006yk}). An
alternative approach is to consider non-linear terms which describe
gluon recombination as in the colour glass condensate
model~\cite{Iancu:2003xm} which has had considerable success in
explaining RHIC data~\cite{JalilianMarian:2005jf}. A recent suggestion
is to use a structure function consistent with HERA data that
saturates the Froissart unitarity bound and thus predicts a $\ln^2 s$
dependence of the cross-section~\cite{Berger:2007ic}. Such approaches
are beyond the scope of the present paper, which is concerned with the
more modest goal of estimating the uncertainties on high energy
neutrino cross-sections which are compatible with the conventional NLO
DGLAP formalism. The motivation is to provide an update on the
neutrino cross-sections in the literature \cite{Gandhi:1998ri} which
are widely used e.g. for estimating event rates in neutrino telescopes
such as Baikal \cite{Antipin:2007zz}, ANTARES \cite{Aslanides:1999vq}
and IceCube \cite{icecube}, cosmic ray observatories such as HiRes
\cite{Martens:2007ff} and Auger \cite{auger}, and radio detectors such
as GLUE \cite{Gorham:2003da}, FORTE \cite{Lehtinen:2003xv}, RICE
\cite{Kravchenko:2002mm} and ANITA \cite{Barwick:2005hn}. As a
corollary, if cross-sections much outside these limits are observed,
it would be a clear signal of the need for extensions to conventional
formalism. To date no unambiguous signals which require such
extensions have been observed. The prospect for measuring the
cross-section using very high energy cosmic neutrinos in order to
distinguish between theoretical suggestions for gluon dynamics at low
$x$ has been discussed by us elsewhere \cite{Anchordoqui:2006ta}.

Previous work on estimating high energy neutrino
cross-sections~\cite{Gandhi:1998ri} used PDF sets which no longer fit
modern data from HERA~\cite{Tung:2004rw} and an {\em ad hoc} procedure
for estimating PDF uncertainties.  The present work improves on this
in several respects.  Firstly, we use a recent PDF analysis which
includes data from all HERA-I running~\cite{Chekanov:2002pv}.
Secondly, we take a consistent approach to PDF uncertainties --- both
model uncertainties and, more importantly, the uncertainties which
derive from the correlated systematic errors of the input data sets
\cite{CooperSarkar:2002yx}.  Thirdly, we use NLO rather than LO
calculations throughout. Fourthly, we use a general-mass variable
flavour number scheme~\cite{Thorne:1997ga,Thorne:2006qt} to treat
heavy quark thresholds.

\section{Formalism}
\label{sec:form}

Parton Density Function (PDF) determinations are global
fits~\cite{Martin:2001es,Martin:2007bv,Pumplin:2002vw,Tung:2006tb,Chekanov:2002pv},
which use inclusive cross-section data and structure function
measurements from deep inelastic lepton hadron scattering (DIS) data.
The kinematics of lepton hadron scattering is described in terms of
the variables $Q^2$, Bjorken $x$, and $y$ which measures the energy
transfer between the lepton and hadron systems.

The double differential charged current (CC) cross-section for
neutrino and antineutrino production on isoscalar nucleon targets are
given by \cite{Devenish:2004pb},
\begin{equation}
 \frac{\mathrm{d}^2\sigma(\nu (\bar{\nu}) N)}{\mathrm{d}x~\mathrm{d}Q^2} = 
 \frac{G_\mathrm{F}^2 M_W^4}{2\pi(Q^2 + M_W^2)^2 x} 
 \sigma_\mathrm{r}(\nu (\bar{\nu}) N)
\end{equation}
where the reduced cross-sections $\sigma_\mathrm{r}(\nu (\bar{\nu})
N)$ are given by
\begin{equation} 
 \sigma_\mathrm{r}(\nu N) = 
 \left[Y_ + F_2^{\nu} (x, Q^2) - y^2 F_\mathrm{L}^{\nu} (x, Q^2) 
 + Y_ - xF_3^{\nu} (x, Q^2) \right],
\end{equation}
and
\begin{equation} 
\sigma_\mathrm{r}(\bar{\nu} N) = \left[Y_ + F_2^{\bar{\nu}}(x, Q^2) - y^2 
 F_\mathrm{L}^{\bar{\nu}}(x, Q^2) - Y_ - xF_3^{\bar{\nu}}(x, Q^2) \right],
\end{equation}
where the structure functions $F_2$, $xF_3$ and $F_\mathrm{L}$ are
related directly to quark momentum distributions.

The QCD predictions for these structure functions are obtained by
solving the DGLAP evolution equations at NLO in the
\mbox{$\overline{\mathrm{MS}}$} scheme with the renormalisation and
factorization scales both chosen to be $Q^2$.  These equations yield
the PDFs at all values of $Q^2$ provided these distributions have been
input as functions of $x$ at some input scale $Q^2_0$.  The resulting
PDFs are then convoluted with coefficient functions, in order to
obtain the structure functions.

We use the PDF fit formalism of the published ZEUS-S global PDF
analysis~\cite{Chekanov:2002pv}, but this fit is updated as
follows. First, the range of the calculation has been extended up to
$Q^2 = 10^{12}$~GeV$^2$ and down to $x = 10^{-12}$.  Second, {\em all}
inclusive cross-section data for neutral and charged current reactions
from ZEUS HERA-I running (1994--2000) are included in the fit.  Third,
the parametrization is extended from 11 to 13 free parameters, input
at $Q^2_0=7$~GeV$^2$. In summary, the PDFs for $u$ valence quarks
($xu_\mathrm{v}(x)$), $d$ valence quarks ($xd_\mathrm{v}(x)$), total
sea quarks ($xS(x)$), and the gluon ($xg(x)$), are each parametrized
by the form
\begin{equation}
  p_1 x^{p_2} (1 - x)^{p_3} P(x),
\label{eqn:pdf}
\end{equation}
where $P(x) = 1 +p_5 x$. The strong coupling constant is taken to be
$\alpha_\mathrm{s}(M_Z^2) = 0.118$~\cite{Eidelman:2004wy}. The total
sea contribution is $xS=2x(\bar{u} +\bar{d} +\bar{s}+ \bar{c}
+\bar{b})$, where $\bar{q}=q_\mathrm{sea}$ for each flavour,
$u=u_\mathrm{v}+u_\mathrm{sea}, d=d_\mathrm{v}+d_\mathrm{sea}$ and
$q=q_\mathrm{sea}$ for all other flavours. The flavour structure of
the light quark sea allows for the violation of the Gottfried sum rule
such that $x(\bar{d}-\bar{u})$ is non-zero, but only the normalisation
of this quantity is free, the shape being fixed in accordance with
E866 Drell-Yan data~\cite{Hawker:1998ty}.  A suppression of the
strange sea with respect to the non-strange sea of a factor of 2 at
$Q^2_0$ is also imposed, consistent with neutrino induced dimuon data
from CCFR~\cite{Bazarko:1994tt}.  The normalisation parameters, $p_1$,
for the $d$ and $u$ valence and for the gluon are constrained to
impose the number sum-rules and momentum sum-rule. The low-$x$ shape
parameters $p_2$ for the $u$ and $d$ valence quarks are set
equal. Finally there are 13 free PDF parameters. Reasonable variations
of these assumptions about the input parametrization are included in
the model uncertainties on the output PDFs.

A more important source of uncertainties on the PDFs comes from the
experimental uncertainties on the input data.  The PDFs are presented
with full accounting for uncertainties from correlated systematic
errors (as well as from statistical and uncorrelated sources) using
the conservative OFFSET method. The uncertainty bands should be
regarded as $68\%$ confidence limits. A full discussion of approaches
to estimating PDF uncertainties is given
in~\cite{Chekanov:2002pv,CooperSarkar:2002yx}. The PDF uncertainties
from this updated ZEUS-S-13 fit are comparable to those on the
published ZEUS-S fit \cite{Chekanov:2002pv}, as well as the most
recent fits of the CTEQ~\cite{Pumplin:2002vw,Tung:2006tb} and
MRST~\cite{Martin:2001es,Martin:2007bv} groups.

Previous work~\cite{Gandhi:1998ri} treated heavy quark production by
using a zero-mass variable flavour number scheme, with slow-rescaling
at the $b$ to $t$ threshold. Although, as explained in
Section~\ref{sec:results}, the exact treatment of the $b \to t$
threshold is not very important for the estimation of high energy
neutrino cross-sections, it is important to use a correct treatment of
heavy quark thresholds when determining the PDFs. We note that the
central values of the sea quark distributions of the most recent
CTEQ6.5 analysis \cite{Tung:2006tb} which uses a general mass variable
flavour scheme for heavy quarks, lie {\em outside} the $90\%$ CL
uncertainty estimates of the previous CTEQ6.1 analysis
\cite{Pumplin:2002vw}, which used a zero-mass variable flavour
number scheme (as did all previous CTEQ analyses). This difference is
significant for lower $Q^2$ ($\lesssim 5000$~GeV$^2$) and middling $x$
($5\times 10^{-5} \lesssim x \lesssim 5\times 10^{-2}$) and this is a
kinematic region of relevance to the present study. The heavy quark
production scheme used in the present fit is the general mass variable
flavour number scheme of Roberts and
Thorne~\cite{Thorne:1997ga,Thorne:2006qt}. The central values of both
the CTEQ6.5 PDF analysis \cite{Tung:2006tb} and the MRST2004 NLO
analysis~\cite{Martin:2004ir} (which also uses a general-mass
variable flavour number scheme) lie within, or very close to, the
uncertainty bands of the present analysis over the entire kinematic
region of interest.

\section{Results}
\label{sec:results}

\begin{figure}[tbp]
\centerline{
\epsfig{figure=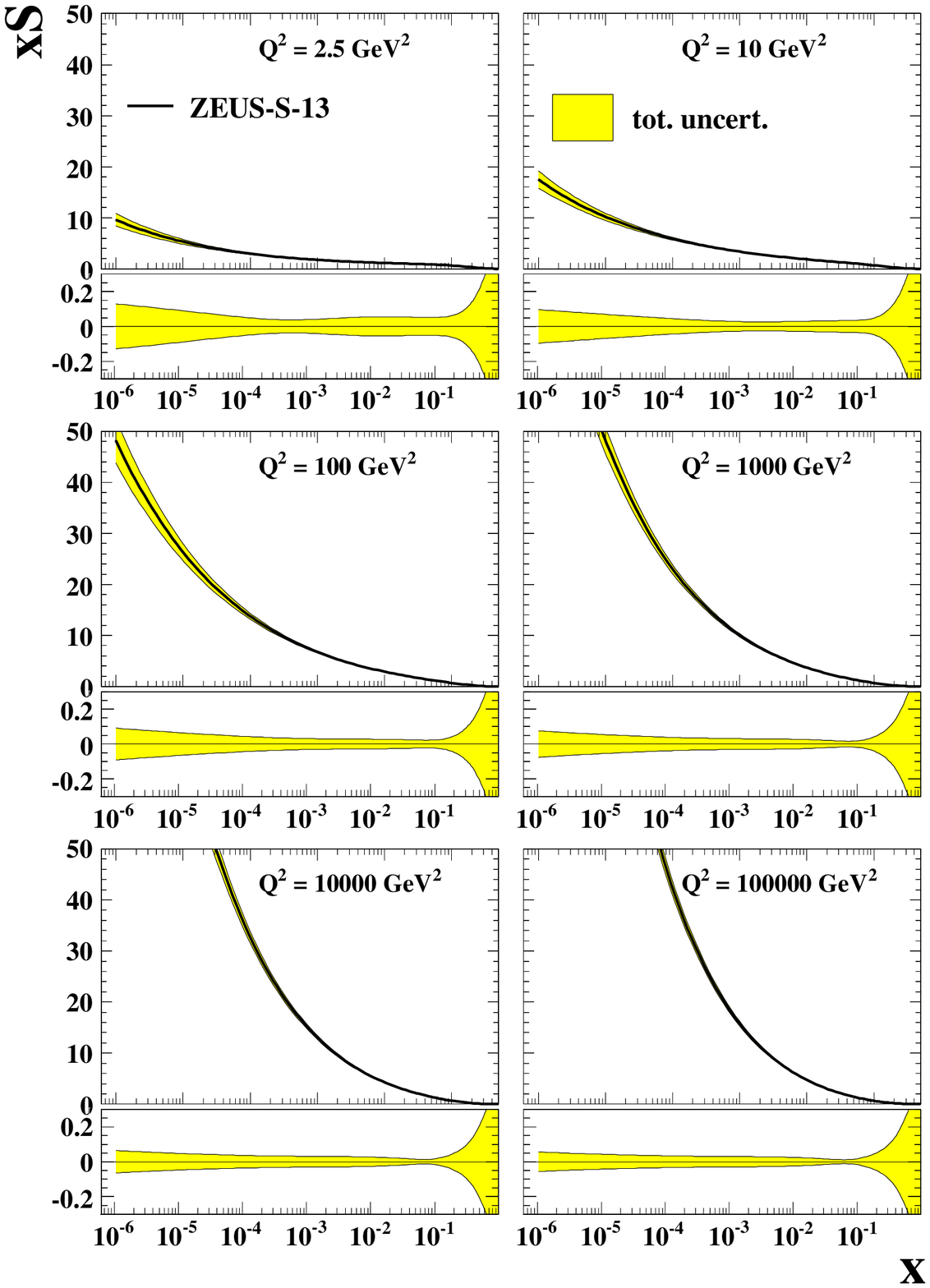,width=0.5\textwidth}
\epsfig{figure=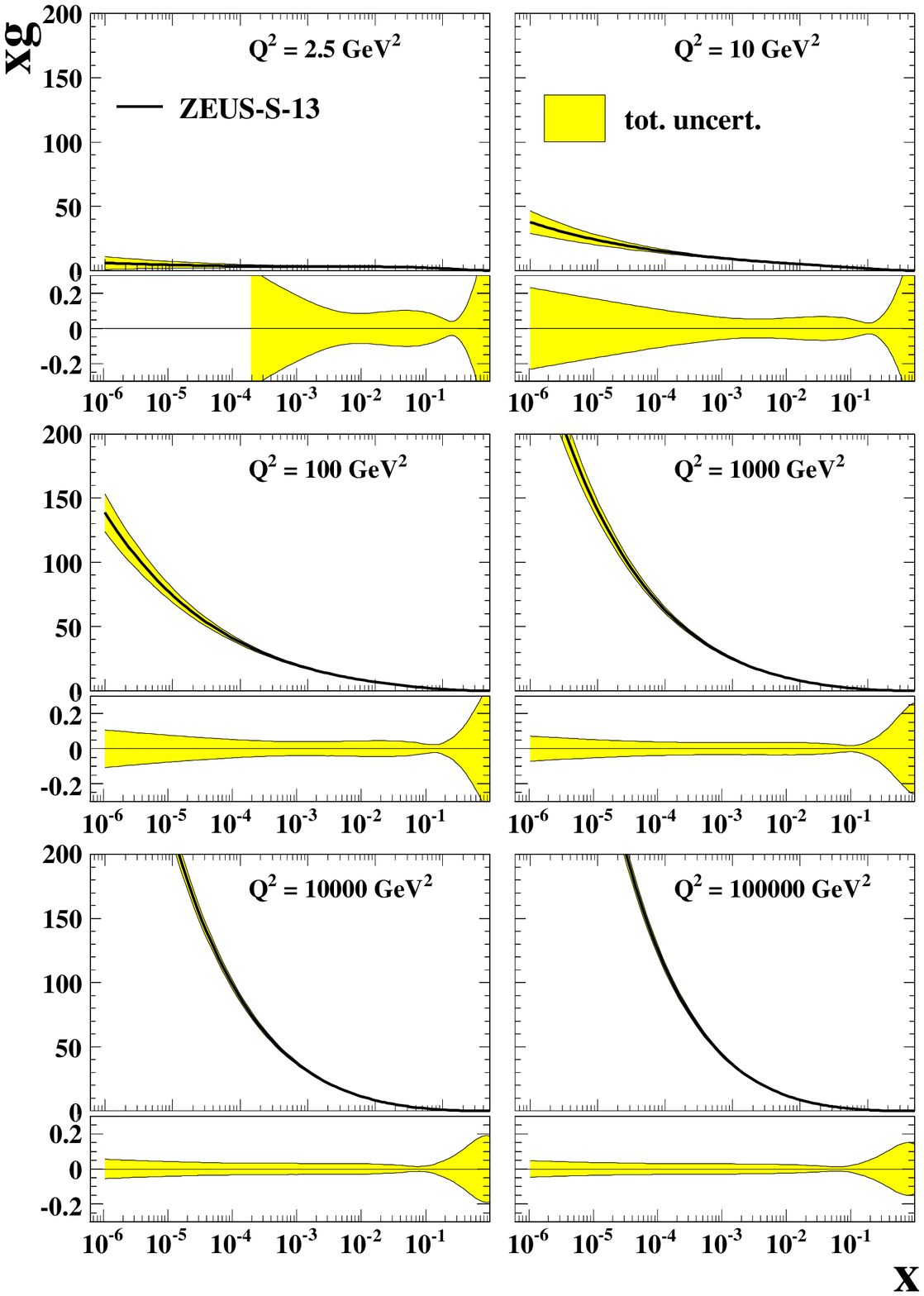,width=0.5\textwidth}}
\caption {The PDFs and their fractional uncertainties at various $Q^2$
 are shown for sea quarks (left) and gluons (right).
}
\label{fig:glusea}
\end{figure}

At very small $x$ and high $Q^2$, the $\nu N$ cross-section is
dominated by sea quarks produced by gluon splitting $g\to q\bar{q}$.
In this kinematic region, the parametrisation of the gluon momentum
distribution is approximately: $xg (x, Q^2) \propto x^{-\lambda}$,
where $\lambda \sim 0.3-0.4$. Figure~\ref{fig:glusea} shows the
predicted sea and gluon distributions from the present PDF fit and
their fractional uncertainties, at various $Q^2$ values. This illustrates
that the PDF uncertainties are largest at low $Q^2$ and at low-$x$.
PDF uncertainties are also large at very high-$x$ but this kinematic
region is not important for scattering of high energy neutrinos.

In QCD at leading order, the longitudinal structure function
$F_\mathrm{L}$ is identically zero, and the structure functions $F_2$
and $xF_3$ for neutrino interactions on isoscalar targets can be
identified with quark distributions as follows:
\begin{equation}
 F_2^{\nu} = x (u + d + 2s + 2b + \bar{u} + \bar{d} + 2\bar{c}),\\ 
 \quad
 xF_3^{\nu} = x (u + d + 2s + 2b - \bar{u} - \bar{d} - 2\bar{c}).
\end{equation}
Similarly for antineutrino interactions,
\begin{equation}
 F_2^{\bar{\nu}} = x (u + d + 2c +\bar{u} + \bar{d} + 2\bar{s} + 2\bar{b}),\\ 
 \quad
 xF_3^{\bar{\nu}} = x (u + d + 2c - \bar{u} - \bar{d} - 2\bar{s} - 2\bar{b}).
\end{equation}
Assuming, $s=\bar{s},\,c=\bar{c},\,b=\bar{b}$, we obtain
$F_2^{\nu}=F_2^{\bar{\nu}}$, whereas
$xF_3^{\nu}-xF_3^{\bar{\nu}}=2(s+\bar{s}+b+\bar{b}-c-\bar{c})=4(s+b-c)$.
At NLO these expressions must be convoluted with appropriate
coefficient functions (such that $F_\mathrm{L}$ is no longer zero) but
these expressions still give us a good idea of the dominant
contributions. Note however that the $b$ contribution will be very
suppressed until $Q^2 \gg M_t^2 \sim 3 \times 10^4$~GeV$^2$, since the
$b\to t$ coupling is dominant.

\begin{figure}[tbp]
\centerline{
\epsfig{figure=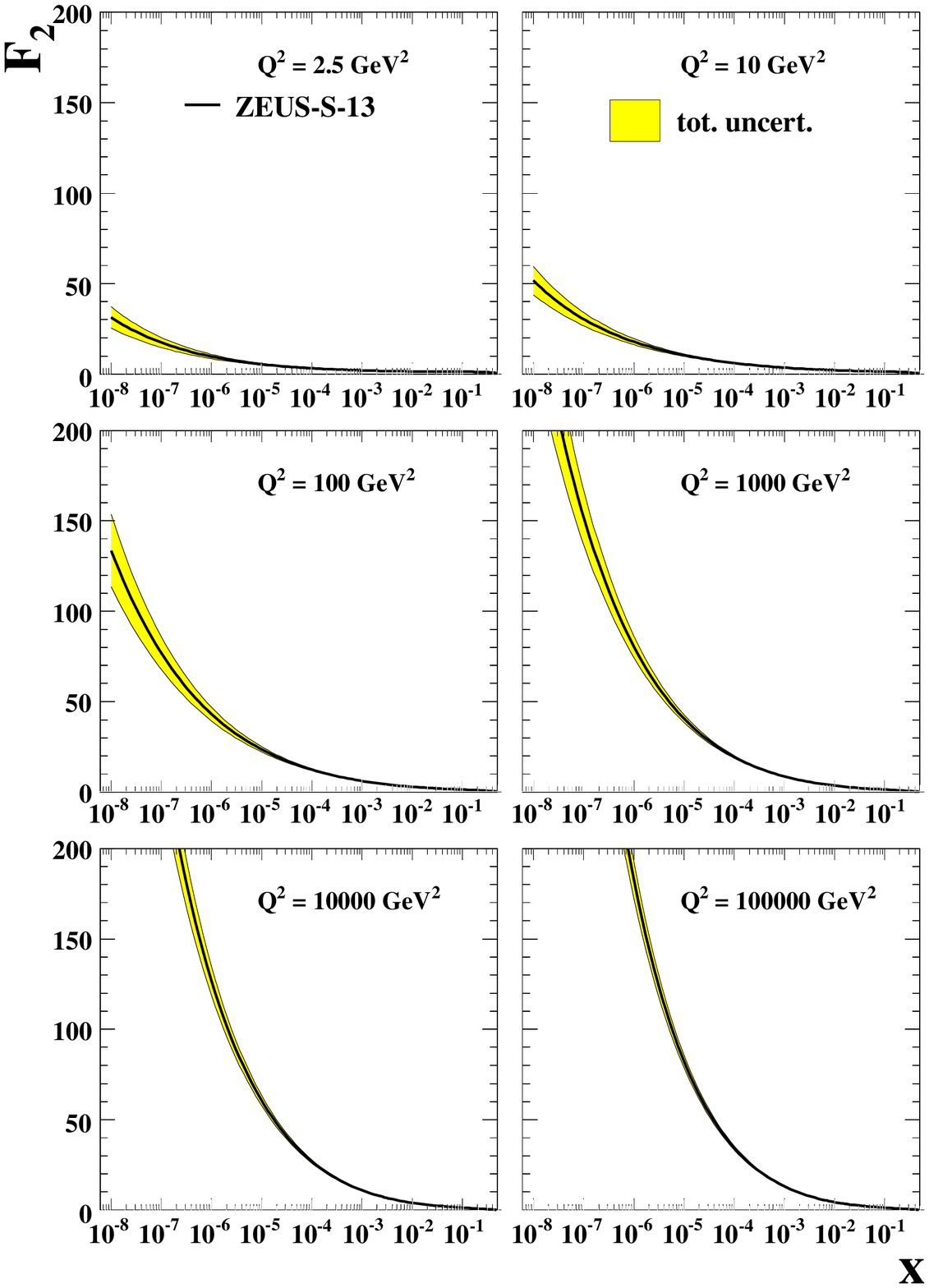,width=0.33\textwidth}
\epsfig{figure=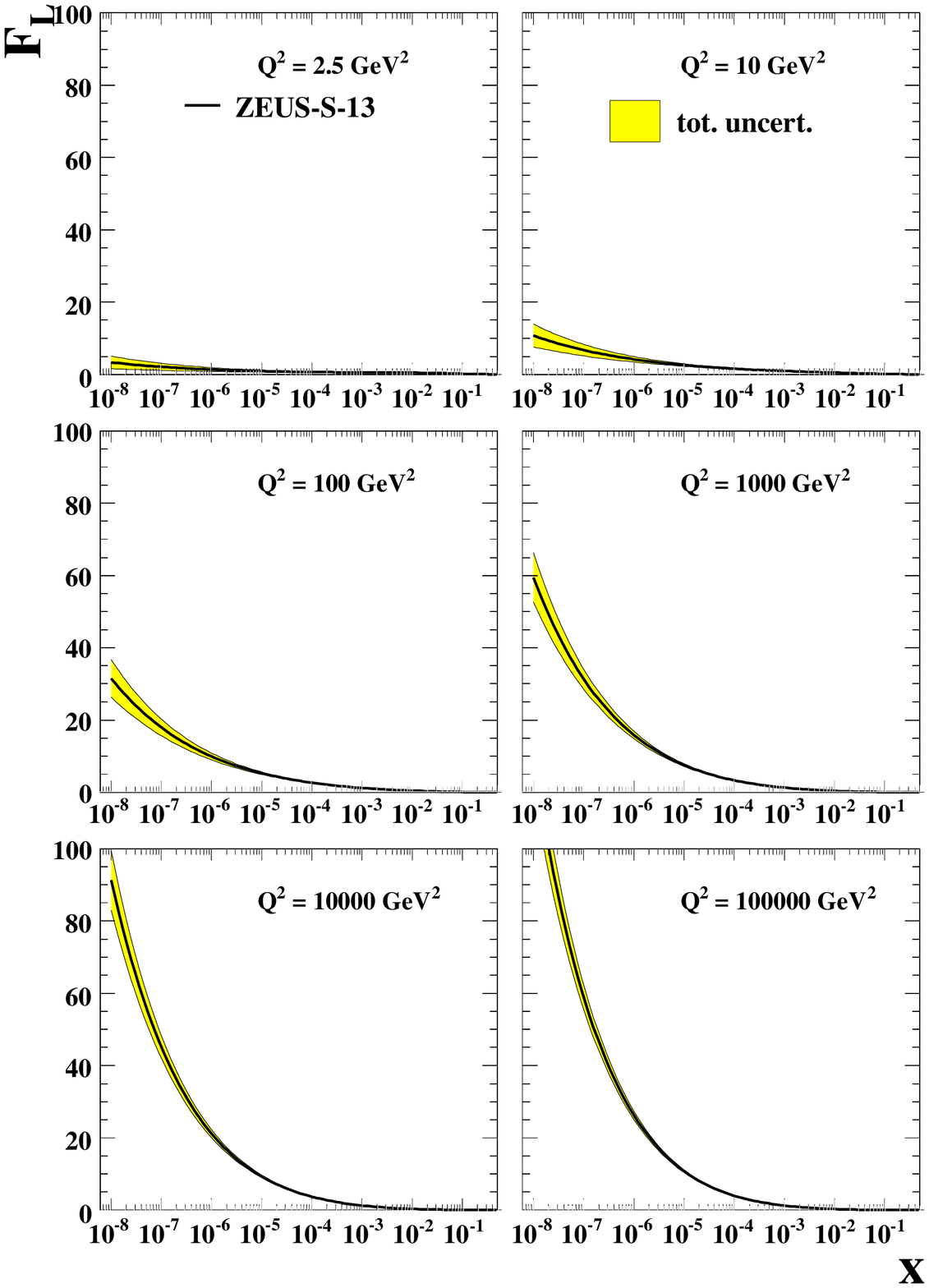,width=0.33\textwidth}
\epsfig{figure=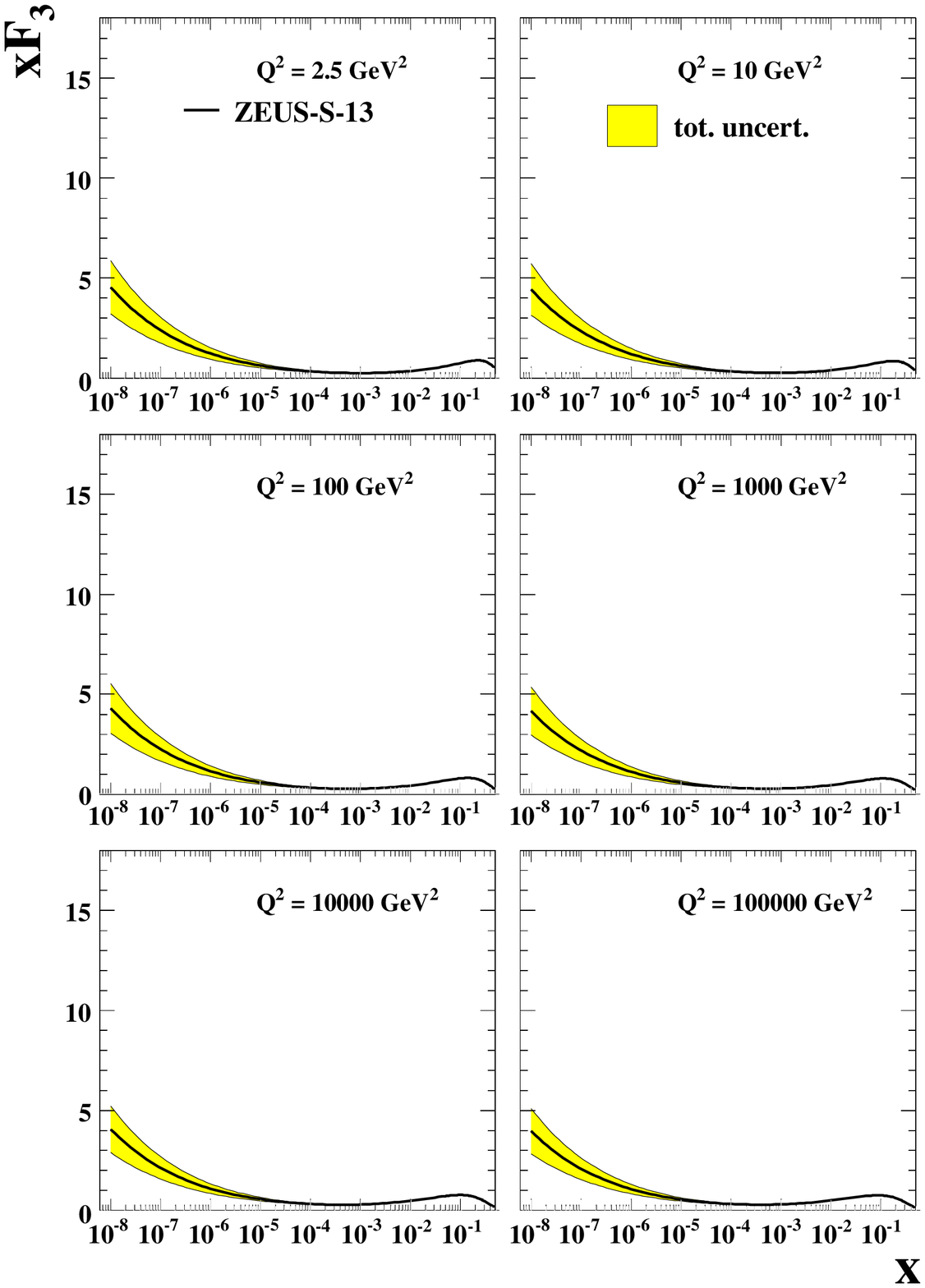,width=0.33\textwidth}}
\centerline{
\epsfig{figure=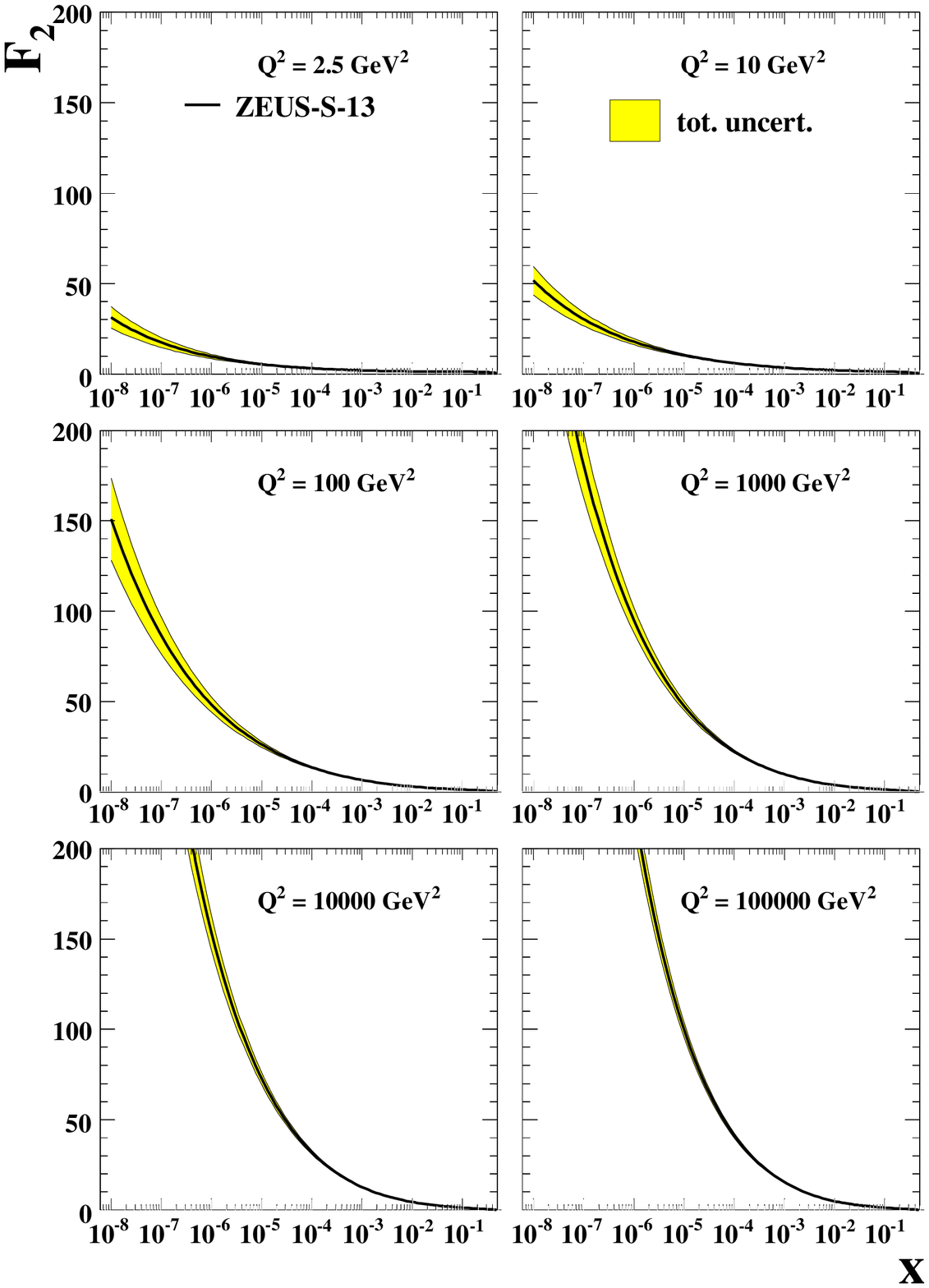,width=0.33\textwidth}
\epsfig{figure=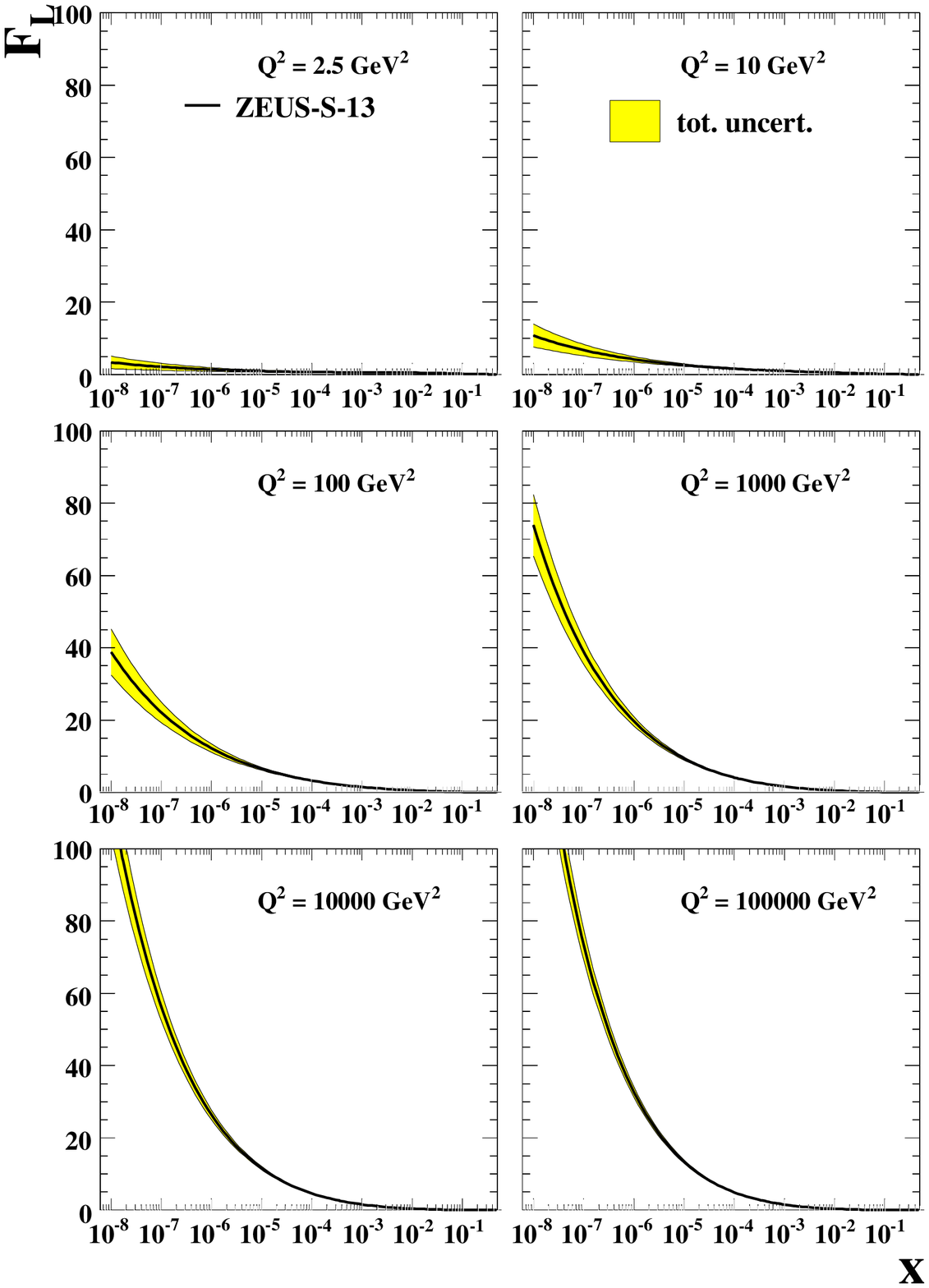,width=0.33\textwidth}
\epsfig{figure=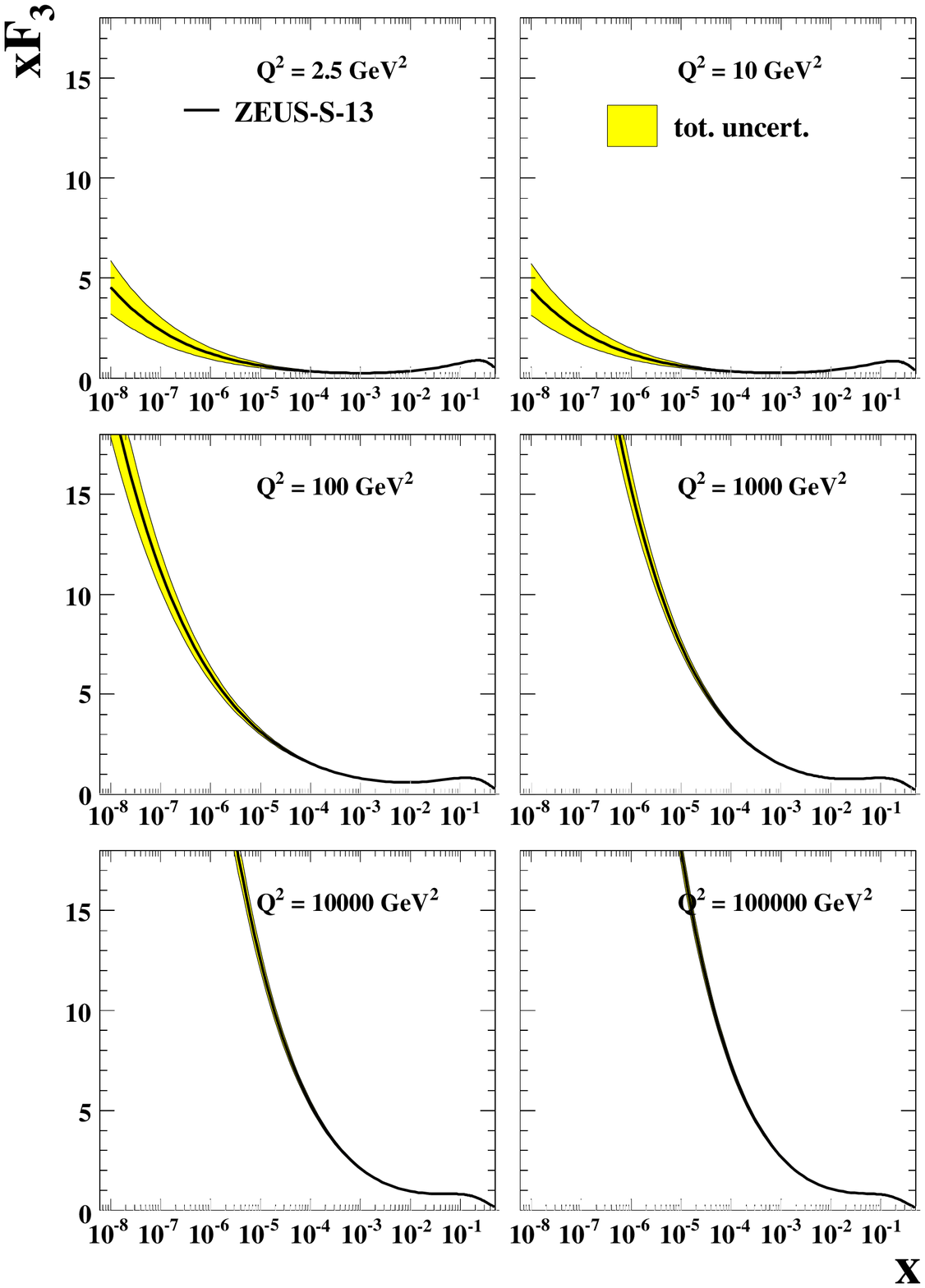,width=0.33\textwidth}
}
\caption {Predictions for $F_2^{\nu}$, $F_L^{\nu}$ and $xF_3^{\nu}$
  using a zero mass variable flavour number scheme, without the $b$
  contribution (upper panels), and with the $b$ contribution (lower
  panels).}
\label{fig:stfns}
\end{figure}

\begin{figure}[tbp]
\centerline{
\epsfig{figure=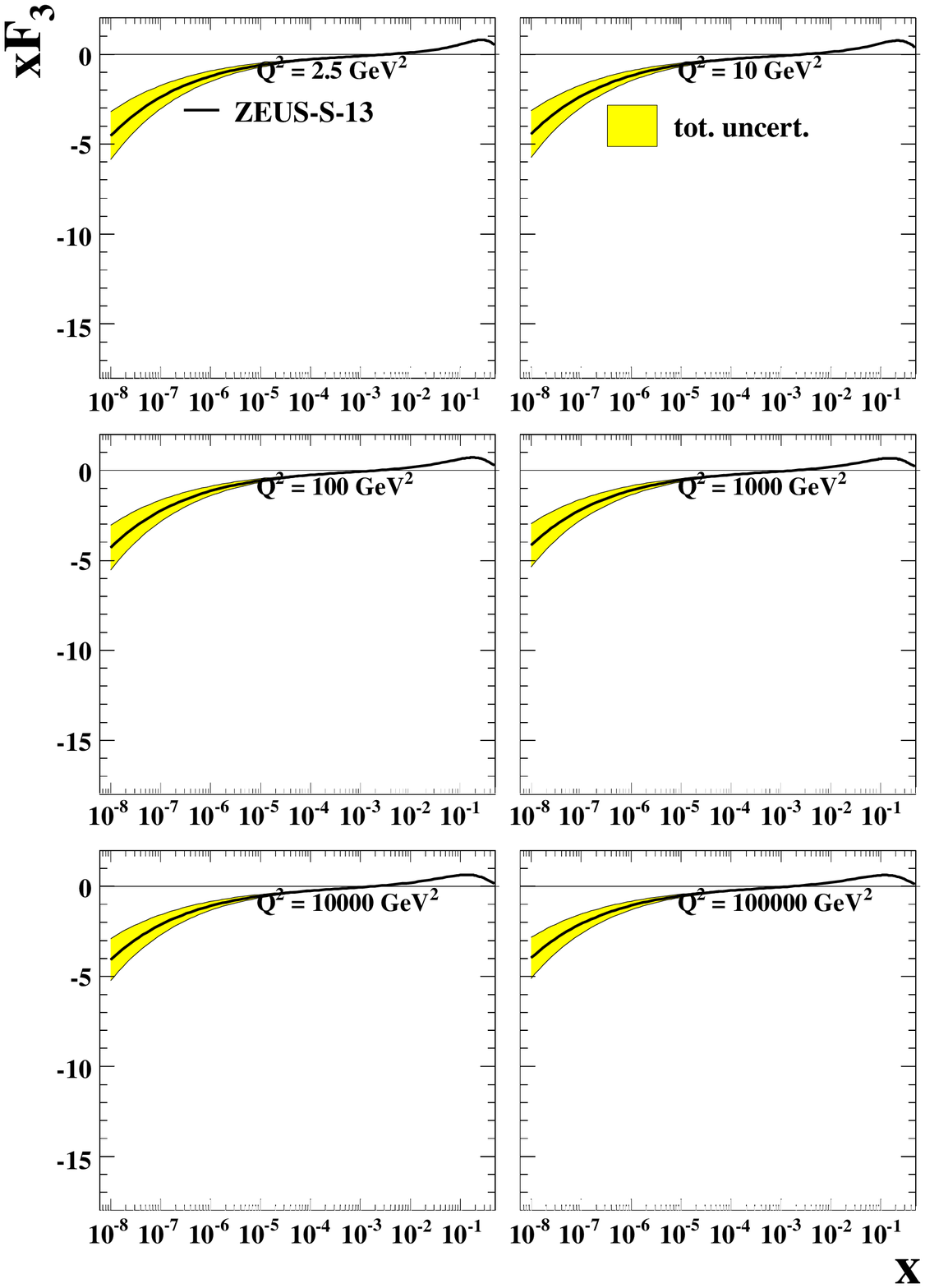,width=0.5\textwidth}
\epsfig{figure=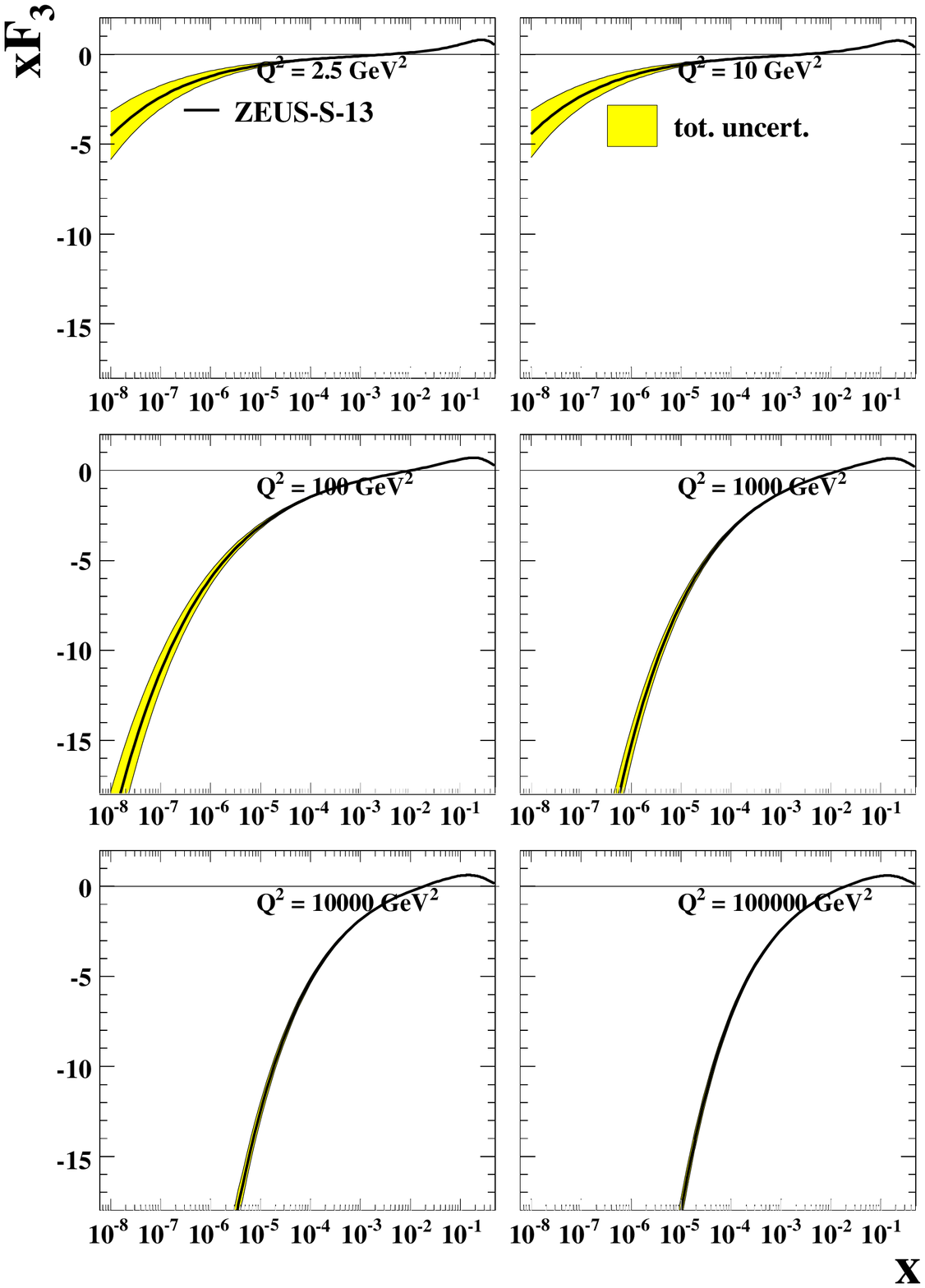,width=0.5\textwidth}
}
\caption{Predictions for $xF_3^{\bar{\nu}}$ for antineutrinos using a
  zero mass variable flavour number scheme, without the $b$
  contribution (left panel), and with the $b$ contribution (right
  panel).}
\label{fig:xf3nubar}
\end{figure}

In Figure~\ref{fig:stfns} we show predictions for the neutrino
structure functions $F_2^{\nu}$, $F_\mathrm{L}^{\nu}$ and $xF_3^{\nu}$
and in Figure~\ref{fig:xf3nubar} we show the antineutrino structure
function $xF_3^{\bar{\nu}}$. In order to illustrate the potential
impact of the $b$ contribution, these were calculated using the
coefficient functions of the zero-mass variable flavour number scheme
with (lower panels) and without (upper panels) the $b$ contribution.
Note that the input PDFs were still determined using the general mass
variable flavour number scheme.

The predictions for $F_2$ and $F_\mathrm{L}$ are somewhat suppressed
without the $b$ contribution, as is expected since the contribution of
$b$ to $F_2$ is at most $20\%$. However, the effect on $xF_3$ is
much more dramatic. This can be understood by considering the LO
expressions
\begin{equation}
 xF_3^{\nu} = x (u_v + d_v + 2(s-\bar{c}) + 2 b)
\end{equation}
and 
\begin{equation}
 xF_3^{\bar{\nu}} = x (u_v + d_v + 2(c-\bar{s}) - 2\bar{b}).
\end{equation}
At low-$x$ the valence contributions are close to zero, while the
strange and charm sea are of opposite sign and nearly equal, such
that $xF_3$ is nearly all $b$ quark and $xF_3^{\nu} \sim
-xF_3^{\bar{\nu}}$.

Even though there are dramatic differences in predictions for $xF_3$
with and without the $b$ contribution, this does not lead to
significant differences in the $\nu N$ and $\bar{\nu} N$
cross-sections because, at low-$x$, $xF_3 \lesssim F_2/5$, and the $y$
dependence suppresses the contribution of $xF_3$ further. The $b$
contribution to the reduced cross-section integrated over $y$ is
always less than $\sim 25\%$. However, in practice it is even more
suppressed for CC processes since $b$ is important only at higher
$Q^2(\gg M_t^2$) because of the $t$ threshold. Furthermore, in the
total cross-section, the $W$ propagator suppresses the contribution of
the kinematic region $Q^2 \gg M_W^2$, such that contributions from
$Q^2 \gg M_t^2$ are suppressed very significantly.

\begin{figure}[tbp]
\centerline{
\epsfig{figure=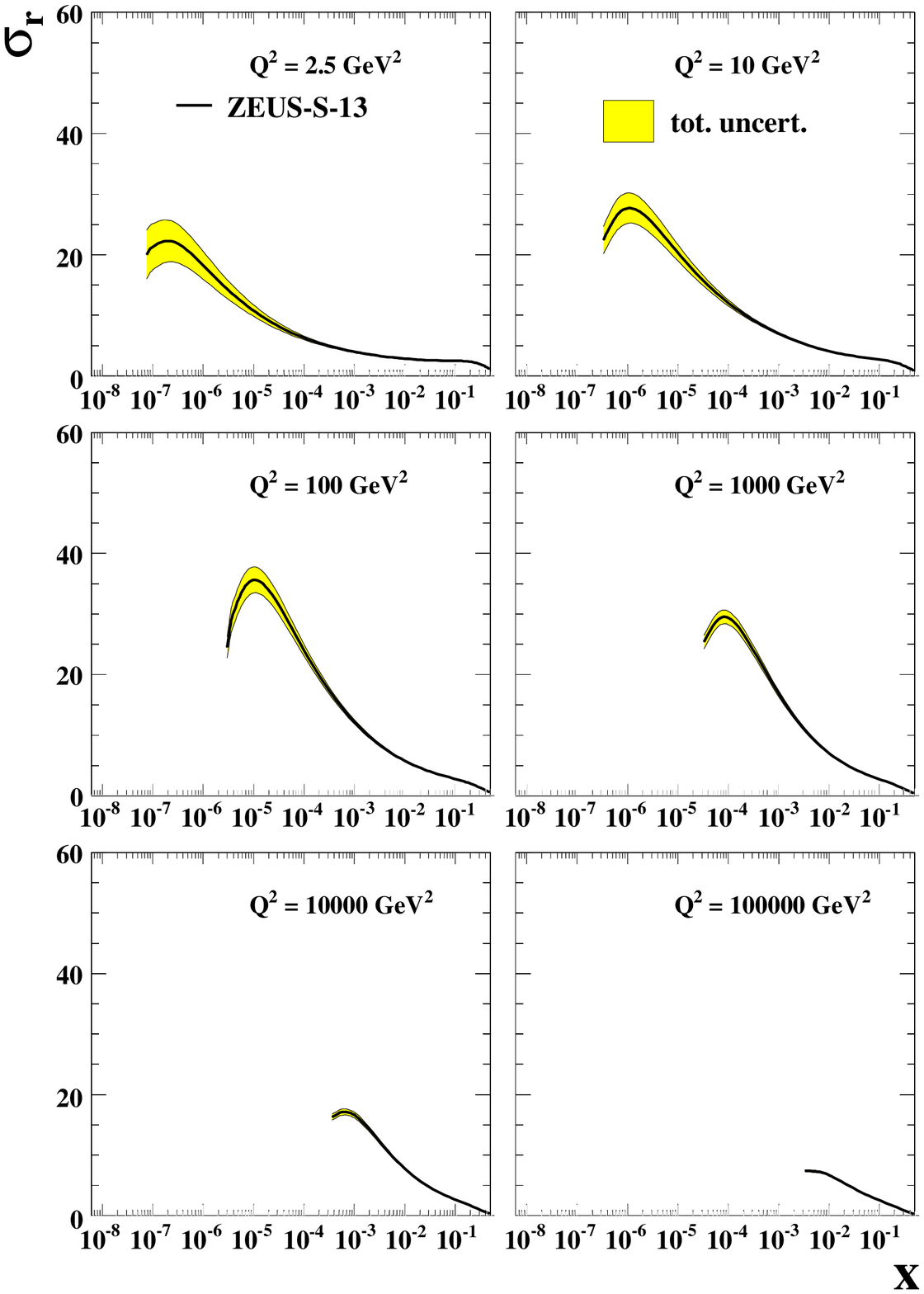,width=0.5\textwidth}
\epsfig{figure=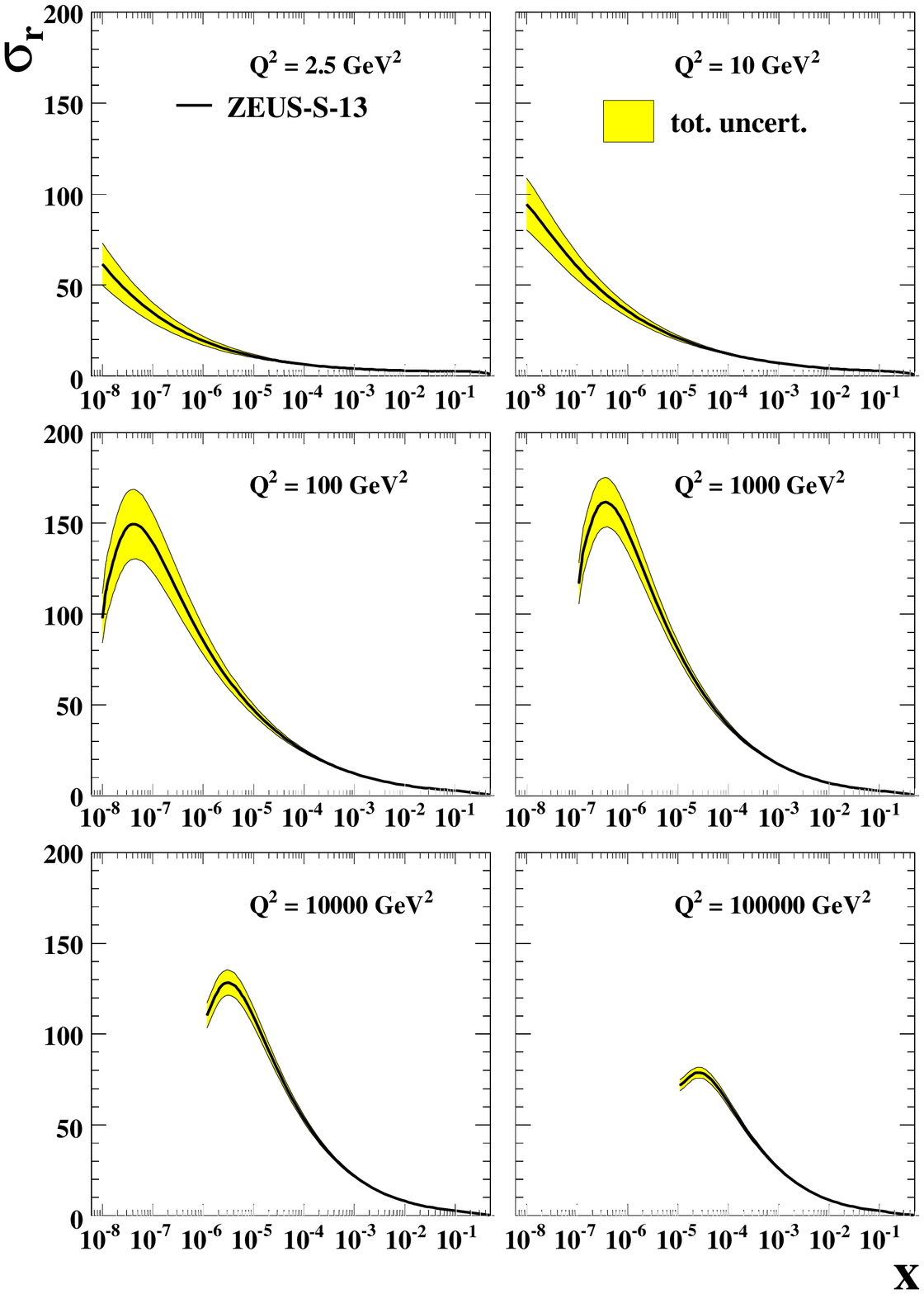,width=0.5\textwidth}}
\caption{Neutrino-nucleon reduced cross-sections for various $Q^2$ at
  $s= 3.6\times 10^7$~GeV$^2$ i.e. $E_\nu=1.9\times 10^7$~GeV (left
  panel), and $s= 10^{10}$~GeV$^2$ i.e. $E_\nu=5.3 \times 10^9$~GeV
  (right panel).}
\label{fig:nuxsecns}
\end{figure}

Figure~\ref{fig:nuxsecns} shows the predictions for the reduced
neutrino cross-sections as a function of $x$ for various $Q^2$ values
above and below $M_W^2$ and $M_t^2$. These illustrations have been
made in terms of the reduced cross-section in order that one can see
how the structure functions (hence the PDFs) contribute to the total
cross-section.  These cross-sections have been calculated using the
coefficient functions of the general-mass variable flavour scheme and
are given for two representative values of the neutrino energy: $s=3.6
\times10^7$~GeV$^2$ ($\Rightarrow E_\nu=1.9 \times 10^7$~GeV) and $s=
10^{10}$~GeV$^2$ ($\Rightarrow E_\nu=5.3 \times 10^9$~GeV). We do not
show the antineutrino cross-sections separately because these are very
close to the neutrino cross-sections at high energy. This is because
the dominant structure function is $F_2^{\nu}=F_2^{\bar{\nu}}$, and
although, $xF_3^{\nu} \sim - xF_3^{\bar{\nu}}$, the structure function
$xF_3$ contributes with {\em opposite} sign in the neutrino and
antineutrino cross-sections such that the net contribution of $xF_3$
is the same.

The restriction on the lowest value of $x$ probed for each $Q^2$ value
is explained by the effect of the kinematic cut-off, $y < 1$: since,
$x=Q^2/sy$, we must have, $x>Q^2/s$. This kinematic cut-off ensures
that higher $Q^2$ values do not probe very low-$x$ until the neutrino
energies are very high indeed. These figures illustrate which regions
of $x$ and $Q^2$ contribute most strongly to the reduced cross-section
for the different neutrino energies. The dominant contributions come
from $50 \lesssim Q^2 \lesssim 10^4$~GeV$^2$ (where the exact region
moves up gradually with $s$); the contribution of higher $Q^2$ ($Q^2 >
M_W^2$) is suppressed by the $W$-propagator. For the lower energy
$E_\nu=1.9\times 10^7$~GeV, the important range is $10^{-6} \lesssim x
\lesssim 10^{-3}$, while for the higher energy $E_\nu=5.3 \times
10^9$~GeV, this moves down to $10^{-8} \lesssim x \lesssim 10^{-4}$.

The PDF uncertainties are large at low-$x$ and low $Q^2$, but since
the dominant contributions to the cross-section do {\em not} come from
very low $Q^2$ values, the PDF uncertainty on the total neutrino
cross-section is quite small even at the highest energies considered
here: $s=10^{12}$~GeV$^2$.

The total neutrino cross-sections are now obtained by integrating the
predicted double differential cross-section
$\mathrm{d}^2\sigma/\mathrm{d}x\mathrm{d}y$ with no cuts on either
kinematic variable.\footnote{Experiments may have specific cuts on
  e.g. $y$ and we are happy to provide the differential cross-section
  for use in simulation programmes.} These cross-sections are
tabulated in Table~\ref{tab:hiexsecns} at various values of $s$
between $10^7$ and $10^{12}$~GeV$^2$, together with their
uncertainties due to the PDFs, including both model uncertainties and
the experimental uncertainties of the input data sets.\footnote{Note
  that these are somewhat smaller than as shown in our previous work
  \cite{Anchordoqui:2006ta} since we have now carefully evaluated the
  effect of heavy quark thresholds on the DGLAP evolution, which had
  been added on previously as a systematic uncertainty.} The trend of
the PDF uncertainties can be understood by noting that as one moves to
higher and higher neutrino energies one also moves to lower and lower
$x$ where the PDF uncertainties are increasing. These energies are
relevant to e.g. the Auger experiment where cosmic neutrinos can be
detected both as quasi-horizontal deeply penetrating air showers and
(specifically $\nu_\tau$s) as Earth-skimming tau showers \cite{auger}.
We have shown elsewhere \cite{Anchordoqui:2006ta} that the ratio of
these two classes of events is a diagnostic of the $\nu N$
cross-section, {\em independently} of the (rather uncertain) cosmic
neutrino flux. The latter determines the absolute rates --- e.g.
assuming that extragalactic sources of the observed ultrahigh energy
cosmic rays generate a neutrino flux saturating the ``Waxman-Bahcall
bound''\cite{Waxman:1998yy}, it would take 10 years of running with a
$3 \times 10^4$~km$^2$ array to tell whether the $\nu N$ cross-section
is suppressed significantly below the (unscreened) Standard Model
prediction. Proposed satellite-borne detectors such as EUSO and OWL
would scan even larger areas and achieve the necessary acceptance
within a few years of running \cite{PalomaresRuiz:2005xw}.

\begin{table}[tbh]
\centering
\begin{tabular}{| c || c | c |}\\
\hline
$s$ [GeV$^2$] & $\sigma(\nu)$ [pb] & PDF uncertainty \\
\hline \hline
$ 10^7$              & 1252  & $\pm 3\%$   \\
$2 \times 10^7$      & 1665  & $\pm 3\%$   \\
$5 \times 10^7$      & 2391  & $\pm 3.5\%$ \\
$10^8$               & 3100  & $\pm 4\%$   \\
$2 \times 10^8$      & 4022  & $\pm 4.5\%$ \\
$5 \times 10^8$      & 5596  & $\pm 5.5\%$ \\
$10^9$               & 7135  & $\pm 6\%$   \\
$2 \times 10^9$      & 9082 & $\pm 6\%$    \\
$5 \times 10^9$      & 12333 & $\pm 6.5\%$ \\
$10^{10}$            & 15456 & $\pm 7\%$   \\
$2 \times 10^{10}$   & 19379 & $\pm 7\%$   \\
$5 \times 10^{10}$   & 25789 & $\pm 8\%$   \\
$10^{11}$            & 31865 & $\pm 8\%$   \\
$2 \times 10^{11}$   & 39434 & $\pm 9\%$   \\
$5 \times 10^{11}$   & 51635 & $\pm 12\%$  \\
$10^{12}$            & 63088 & $\pm 14\%$  \\
\hline
\end{tabular}
\caption{Neutrino-nucleon total CC cross-section, with the 
  associated PDF uncertainty, at high energies.}
\label{tab:hiexsecns}
\end{table}

\begin{figure}[tbp]
\centerline{
\epsfig{figure=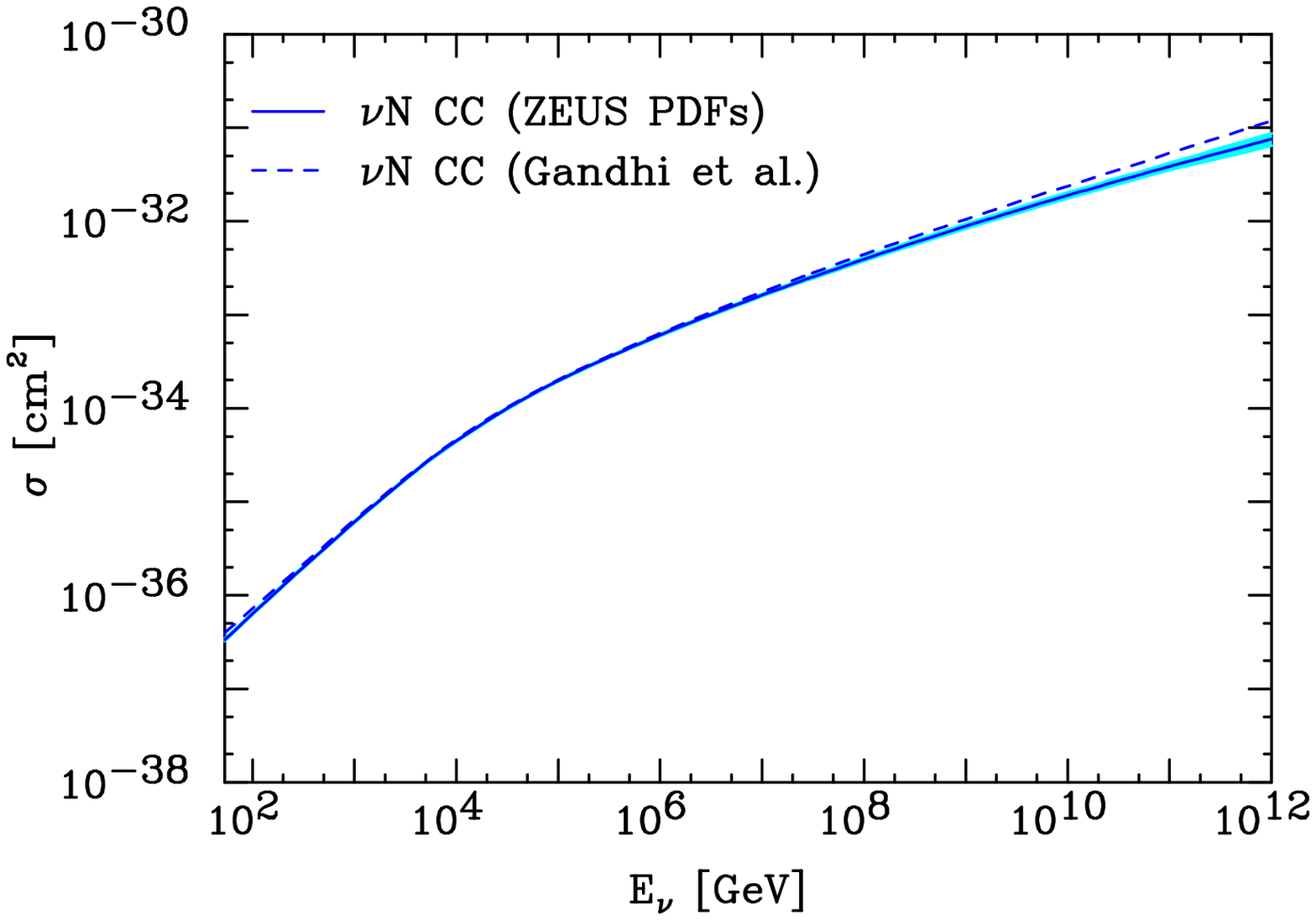,angle=90,width=0.5\textwidth}
\epsfig{figure=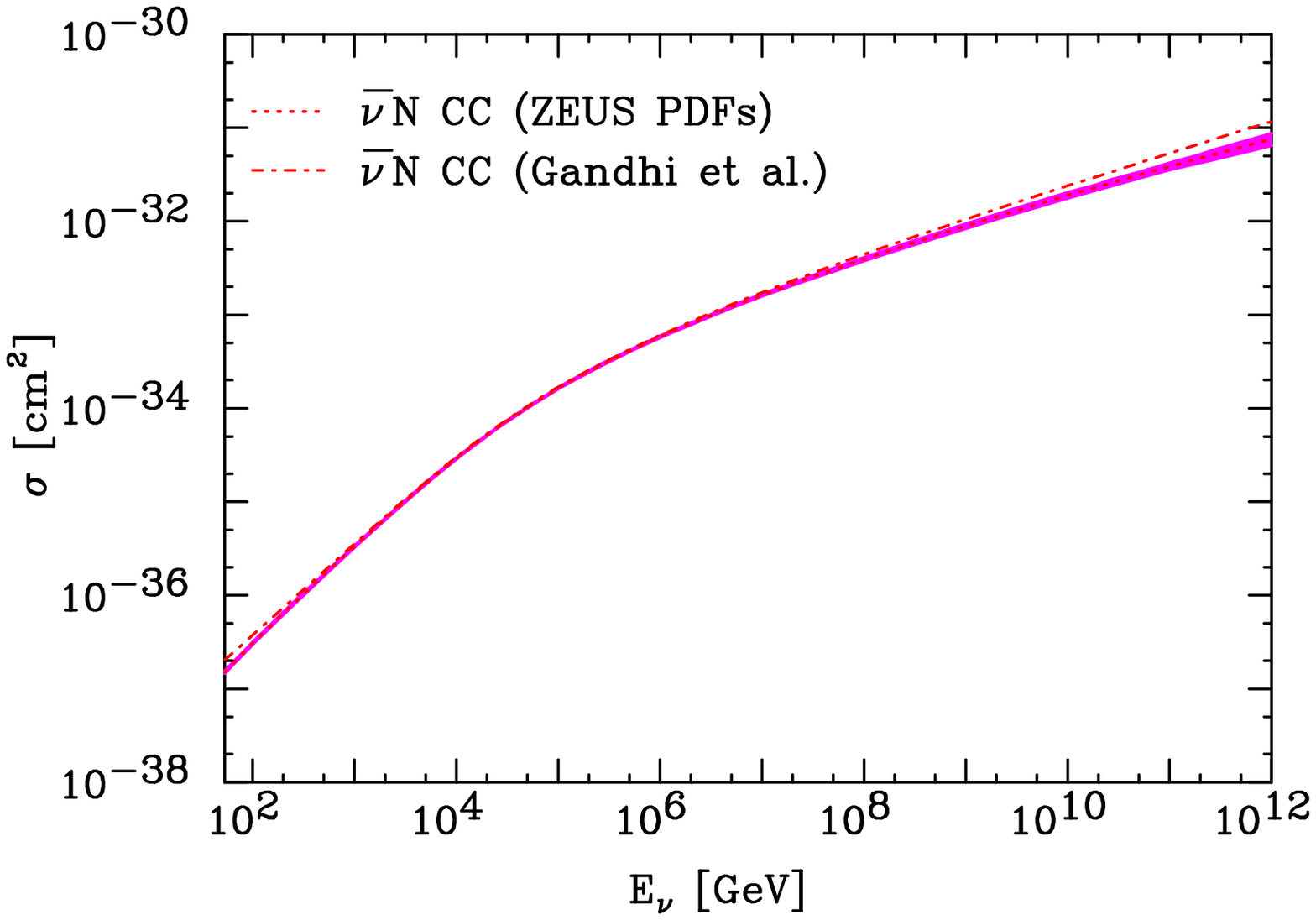,angle=90,width=0.5\textwidth}}
\caption {The total CC cross-section at ultra high energies for
  neutrinos (left) and antineutrinos (right) along with the $\pm
  1\sigma$ uncertainties (shaded band), compared with the previous
  calculation by Gandhi {\it et al}~\cite{Gandhi:1998ri}.}
\label{fig:comparison}
\end{figure}

Figure~\ref{fig:comparison} compares our CC cross-section to the
widely used leading-order calculation of Gandhi {\it et
  al}~\cite{Gandhi:1998ri} which they fitted as:
($\sigma^\mathrm{LO}_\mathrm{CC}/\mathrm{pb}) = 5.53
(E_\nu/\mathrm{GeV})^{0.363}$ for $10^7 \leq (E_\nu/\mathrm{GeV}) \leq
10^{12}$. The present results show a less steep rise of the
cross-section at high energies, reflecting the fact that more recent
HERA cross-section data display a less dramatic rise at low-$x$ than
the early data which was used to calculate the CTEQ4-DIS PDFs. A
power-law description is no longer appropriate over the whole range
$10^7 \leq (E_\nu/\mathrm{GeV}) \leq 10^{12}$, instead the relation
\begin{equation}
  \ln\left(\frac{\sigma^\mathrm{NLO}_\mathrm{CC}}{\mathrm{pb}}\right) = 
\ln(10^{36}) 
- 98.8 \left[\ln\left(\frac{E_\nu}{\mathrm{GeV}}\right)\right]^{-0.0964} ,
\label{fit}
\end{equation}
fits the calculated cross-section to within $\sim10\%$ (P. Mertsch,
private communication).

\begin{figure}[tbp]
\centerline{
\epsfig{figure=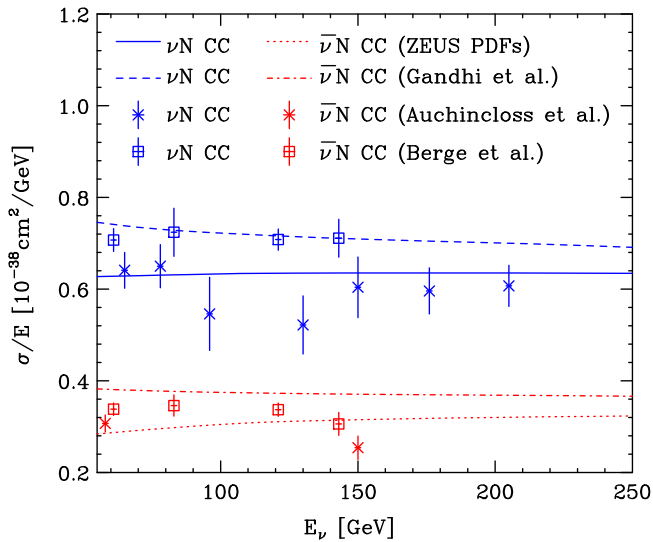,angle=90,width=0.5\textwidth}
}
\caption {The total CC cross-sections at medium energies for neutrinos
  and antineutrinos
 compared with with the previous calculation by Gandhi {\it et
    al}~\cite{Gandhi:1998ri} and with selected experimental data.}
\label{fig:comparisonexpt}
\end{figure}

Neutrino telescopes such as IceCube are optimised to probe lower
energies of order a TeV \cite{icecube}. In this case the high-$x$
region becomes important and the neutrino and antineutrino
cross-sections are {\em different} because the valence contribution to
$xF_3$ is now significant. In Table~\ref{tab:loexsecns} we give both
neutrino and antineutrino cross-sections for values of $s$ between
$10^2$ and $10^7$~GeV$^2$, together with estimates of their PDF
uncertainties. The onset of the linear dependence of the
cross-section on $s$ for $s < M_W^2$ can be seen and in
Figure~\ref{fig:comparisonexpt} we compare the calculated
$\sigma/E_\nu$ with some available recent experimental measurements
\cite{Auchincloss:1990tu,Berge:1987zw}, from the compendium on the
Durham-HEPDATA database \cite{Whalley:2004sz}. The agreement is quite
good, given in particular that our predictions are made for
$Q^2>1$~GeV$^2$ (since perturbative QCD cannot be used at lower
values); for $s \lesssim 100$~GeV$^2$, there can be contributions of
${\cal O}(10\%)$ from even lower values of $Q^2$ which are {\em not}
accounted for here. However there are also corrections for nuclear
shadowing which ought to suppress the cross-section by a comparable
amount. A recent discussion of such effects can be found in
ref.\cite{Armesto:2007tg}.

The trend of the PDF uncertainties can be understood as follows: as
one moves to lower neutrino energies one moves out of the very low-$x$
region such that PDF uncertainties decrease. These uncertainties are
smallest at $ 10^{-2} \lesssim x \lesssim 10^{-1}$, corresponding to
$s\sim 10^5$. Moving to yet lower neutrino energies brings us into the
high-$x$ region where PDF uncertainties are larger again.

\begin{table}[tbh]
\centering
\begin{tabular}{| c || c | c || c | c |}\\
\hline
$s$ [GeV$^2$] & $\sigma(\nu)$ [pb] & PDF uncertainty 
& $\sigma(\bar\nu)$ [pb] & PDF uncertainty\\
\hline\hline
$10^2$           & 0.334 & $\pm 3\%$   & 0.151 & $\pm 4\%$   \\ 
$2 \times 10^2$  & 0.676 & $\pm 2.5\%$ & 0.327 & $\pm 3.5\%$ \\ 
$5 \times 10^2$  & 1.69  & $\pm 2.5\%$ & 0.864 & $\pm 3.5\%$ \\ 
$10^3$           & 3.32  & $\pm 2\%$   & 1.76  & $\pm 3\%$   \\
$2 \times 10^3$  & 6.47  & $\pm 2\%$   & 3.55  & $\pm 2.5\%$ \\
$5 \times 10^3$  & 15.0  & $\pm 2\%$   & 8.67  & $\pm 2.5\%$ \\
$10^4$           & 27.6  & $\pm 2\%$   & 16.6  & $\pm 2.5\%$ \\
$2 \times 10^4$  & 47.0  & $\pm 2\%$   & 30.8  & $\pm 2\%$   \\
$5 \times 10^4$  & 89.4  & $\pm 2\%$   & 64.8  & $\pm 2\%$   \\
$10^5$           & 138   & $\pm 1.5\%$ & 107   & $\pm 1.5\%$ \\
$2 \times 10^5$  & 204   & $\pm 2\%$   & 171   & $\pm 2\%$   \\
$5 \times 10^5$  & 326   & $\pm 2\%$   & 293   & $\pm 2\%$   \\
$10^6$           & 454   & $\pm 2\%$   & 423   & $\pm 2\%$   \\
$2 \times 10^6$  & 628   & $\pm 2.5\%$ & 600   & $\pm 2.5\%$ \\
$5 \times 10^6$  & 937   & $\pm 2.5\%$ & 915   & $\pm 2.5\%$ \\
\hline
\end{tabular}
\caption{Total CC cross-section for neutrinos and antineutrinos with their 
  associated uncertainties at medium energies.}
\label{tab:loexsecns}
\end{table}

\section{Conclusions}
\label{sec:conc}

We have calculated the charged current neutrino cross-section at NLO
in the Standard Model using the best available DIS data along with a
careful estimate of the associated uncertainties.  As mentioned
earlier, there are further uncertainties associated with QCD effects
at very low $x$ which are not addressed in the DGLAP formalism. When
$x$ is sufficiently small that $\alpha_\mathrm{s}\,\ln (1/x) \sim 1$,
it is necessary to resum these large logarithms using the BFKL
formalism. Whereas such calculations at leading-log suggest an even
steeper rise of the gluon structure function at low $x$ (which would
imply a higher $\nu-N$ cross-section), this rise is not so dramatic at
next-to-leading-log; for a recent application of NLL BFKL resummation
to deep inelastic scattering see~\cite{White:2006yh}.  Moreover both
the DGLAP and the BFKL formalisms neglect non-linear screening effects
due to gluon recombination which may lead to saturation of the gluon
structure function. This has been modelled in the colour dipole
framework in which DIS at low $x$ is viewed as the interaction of the
$q \bar q$ dipole to which the gauge bosons fluctuate. An unified
BFKL/DGLAP calculation \cite{Kwiecinski:1998yf} supplemented by
estimates of screening and nuclear shadowing effects, predicts a {\em
  decrease} of the $\nu-N$ cross-section by $20-100\%$ at very high
energies $E_\nu \sim 10^8-10^{12}$~GeV~\cite{Kutak:2003bd}.  An
alternative approach uses the colour glass condensate
formalism~\cite{Iancu:2003xm} and predicts a similar suppression when
a dipole model~\cite{Kharzeev:2004yx} which fits data from RHIC is
used~\cite{Henley:2005ms}. The predicted cross-section is even
lower~\cite{Henley:2005ms} if a different dipole
model~\cite{Bartels:2002cj} developed to fit the HERA data is used and
the gluon distribution is assumed to decrease for $x < 10^{-5}$.
Other possibilities for the behaviour of the high energy $\nu-N$
cross-section have also been
discussed~\cite{Berger:2007ic,Jalilian-Marian:2003wf}.

Detectors for UHE cosmic neutrinos would be able to probe such new
physics if they can establish deviations from the perturbative DGLAP
prediction. Hence we recommend our calculated values for estimation of
the baseline event rates in neutrino telescopes and for use in event
generators such as ANIS \cite{Gazizov:2004va}.  While the expected
neutrino fluxes (e.g. from the sources of the observed high energy
cosmic rays) are rather uncertain, experiments can in principle
exploit the different dependence on the cross-section of the rate of
Earth-skimming and quasi-horizontal events \cite{Anchordoqui:2006ta}.

\acknowledgments{We thank Robert Thorne for discussions and Claire
  Gwenlan and Philipp Mertsch for help with the figures. SS
  acknowledges a PPARC Senior Fellowship (PPA/C506205/1) and the EU
  network ``UniverseNet'' (MRTN-CT-2006-035863); he wishes to thank
  colleagues in Auger and IceCube for encouragement to publish this
  study.}

\end{document}